\documentclass[runningheads]{llncs}
\usepackage[T1]{fontenc}

\usepackage{graphicx}
\usepackage{color}
\usepackage{amsmath}
\usepackage{multirow}
\usepackage{multicol}
\usepackage{mathtools}
\usepackage{mathrsfs}

\usepackage{amsmath,amssymb,amsfonts}
\usepackage{algorithmic}
\usepackage{graphicx}
\usepackage{textcomp}
\usepackage{algorithm}
\usepackage{graphics}
\usepackage{threeparttable}
\usepackage{color}
\usepackage[normalem]{ulem}
\usepackage{multirow}
\usepackage{float}
\usepackage{amsfonts}
\usepackage{bm}
\usepackage{array}
\usepackage{colortbl}
\usepackage{pifont}
\usepackage{diagbox}
\usepackage{rotating}
\usepackage{booktabs}
\usepackage{overpic}
\usepackage{makecell}
\usepackage{contour}
\usepackage[misc]{ifsym} 
\usepackage{hyperref}
\hypersetup{
    colorlinks=true,
    linkcolor=blue,
    filecolor=magenta,      
    urlcolor=blue,
    pdftitle={Overleaf Example},
    pdfpagemode=FullScreen,
    }

\def\ourmodel{CU-Reg}
\begin{document}

\newcommand{\tabref}[1]{Table \ref{#1}}
\newcommand{\tickYes}{\bullet}
\newcommand{\cmark}{\ding{51}}
\newcommand{\tickNo}{\hspace{1pt}\ding{55}}
\newcommand{\figref}[1]{Fig. \ref{#1}}
\newcommand{\supp}[1]{\textcolor{magenta}{#1}}
\newcommand{\notsure}[1]{{\textcolor{red}{#1}}}
\newcommand{\secref}[1]{Sec. \ref{#1}}
\newcommand{\sArt}{state-of-the-art}

\newcommand{\pjl}[1]{{\textcolor{red}{#1}}}

\newcommand{\ie}{{\emph{i.e.}}}
\newcommand{\eg}{{\emph{e.g.}}}
\newcommand{\etal}{{\emph{et al.}}}

\title{Epicardium Prompt-guided Real-time Cardiac Ultrasound Frame-to-volume Registration}
\titlerunning{Real-time Cardiac Ultrasound Frame-to-volume Registration}

\author{Long Lei\thanks{Equal contribution.}\inst{1}\and
Jun Zhou\inst{\star}\inst{2}\and
Jialun Pei\inst{1(\textrm{\Letter})} \and
Baoliang Zhao\inst{3}\and
Yueming Jin\inst{4} \and
Yuen-Chun Jeremy Teoh\inst{1} \and
Jing Qin\inst{2}\and
Pheng-Ann Heng\inst{1}}

\authorrunning{L. Lei et al.}

\institute{The Chinese University of Hong Kong, Hong Kong, China\and
The Hong Kong Polytechnic University, Hong Kong, China\and
Shenzhen Institute of Advanced Technology, CAS, Shenzhen, China\and
National University of Singapore, Singapore, Singapore\\
\email{jialunpei@cuhk.edu.hk}}

\maketitle         
\begin{abstract}

Real-time fusion of intraoperative 2D ultrasound images and the preoperative 3D ultrasound volume based on the frame-to-volume registration can provide a comprehensive guidance view for cardiac interventional surgery.
However, cardiac ultrasound images are characterized by a low signal-to-noise ratio and small differences between adjacent frames, coupled with significant dimension variations between 2D frames and 3D volumes to be registered, resulting in real-time and accurate cardiac ultrasound frame-to-volume registration being a very challenging task. 
This paper introduces a lightweight end-to-end \textbf{C}ardiac \textbf{U}ltrasound frame-to-volume \textbf{Reg}istration network, termed \textbf{CU-Reg}.
Specifically, the proposed model leverages epicardium prompt-guided anatomical clues to reinforce the interaction of 2D sparse and 3D dense features, followed by a voxel-wise local-global aggregation of enhanced features, thereby boosting the cross-dimensional matching effectiveness of low-quality ultrasound modalities. 
We further embed an inter-frame discriminative regularization term within the hybrid supervised learning to increase the distinction between adjacent slices in the same ultrasound volume to ensure registration stability. 
Experimental results on the reprocessed CAMUS dataset demonstrate that our CU-Reg surpasses existing methods in terms of registration accuracy and efficiency, meeting the guidance requirements of clinical cardiac interventional surgery. 
Our code is available at \url{https://github.com/LLEIHIT/CU-Reg}.


\end{abstract}

\section{Introduction}
Cardiac interventional surgery has been widely used in the treatment of structural heart diseases, such as congenital heart disease and valvular heart disease \cite{al2022future}. Compared to DSA (Digital Subtraction Angiography) and CT, 2D ultrasound imaging has the advantages of low equipment requirements, easy operation, real-time imaging, and no radiation exposure, so ultrasound-guided cardiac interventional surgery has become a new trend \cite{abbas2020basic,liu2023automated}. However, 2D ultrasound imaging can only display one section of the heart at a time. Doctors need to determine the position of the section in the heart structure reconstructed in their mind, and further fusion the real-time ultrasound images with the virtual cardiac anatomy to guide the surgical instruments \cite{hacihaliloglu2020interventional}, which requires extremely high levels of doctor experience. 
Currently, 3D ultrasound imaging is also becoming increasingly popular to obtain the complete anatomical structure of the heart \cite{avola2021ultrasound,song2021cross}. 
To provide a complete guidance view for cardiac interventions, it is necessary to explore frame-to-volume registration that fuse intraoperative 2D ultrasound images and preoperative 3D ultrasound volumes in real time, which shortens the learning curve of ultrasound-guided cardiac interventions.

\begin{figure*}[t]
    \centering
    \includegraphics[width=0.98\linewidth]{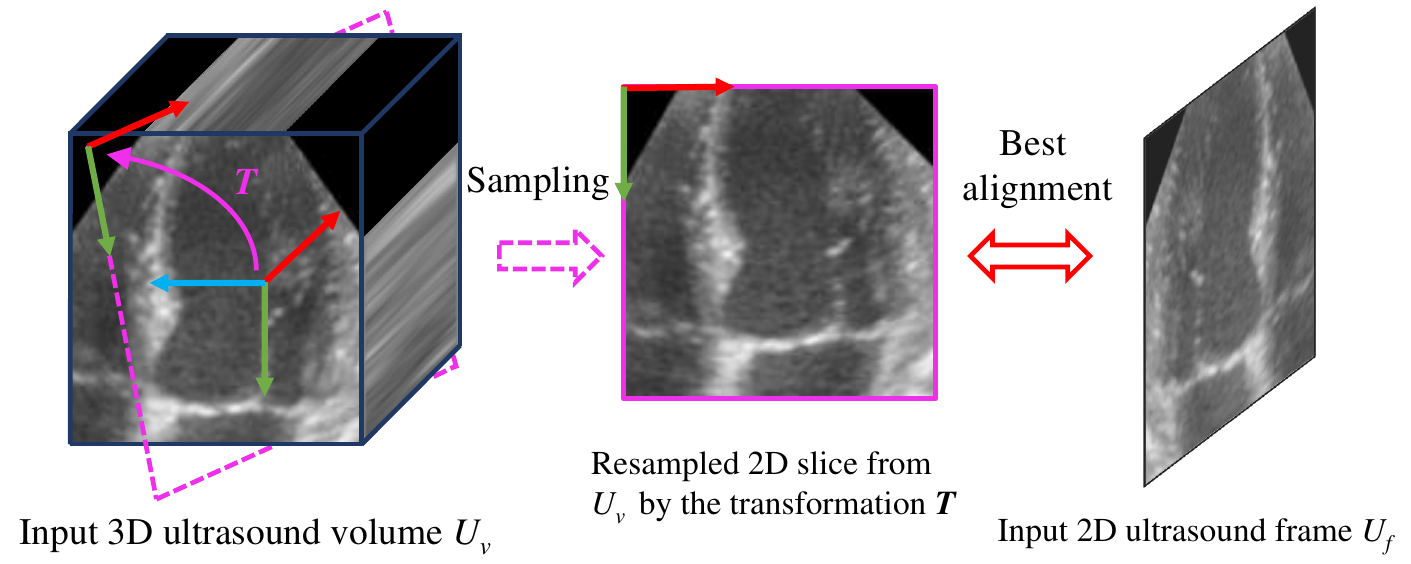}
    \caption{Schematic of cardiac ultrasound frame-to-volume registration.}
    \label{fig:reg_def}
\end{figure*}

The ultrasound frame-to-volume registration aims to seek a transformation that optimally aligns the resampled slice from the given volume by the transformation with the 2D input image~\cite{guo2021end,ferrante2017slice}, as shown in \figref{fig:reg_def}. 
Existing registration methods are divided into mathematical methods and deep learning-based methods. Mathematically, the registration task is usually modeled as an optimization problem~\cite{porchetto2017rigid,lei2023robotic}. 
Although iteration-based methods can yield reasonable accuracy, they cannot meet the real-time requirements of cardiac surgical guidance due to the slow registration speed.
Currently, various deep learning-based methods are widely applied to the image registration task, such as directly learning target transformations~\cite{xu2022svort,bharati2022deep}, keypoint descriptors~\cite{markova2022global}, and image similarity metrics~\cite{haskins2020deep}.
In the field of frame-to-volume registration, Hou~\etal~\cite{hou2017predicting} utilized a CNN-based model to predict rigid transformation of arbitrary 2D image slices from 3D volumes, but only attained an average alignment error of 7 mm on simulated MRI brain data. 
Yeung~\etal~\cite{yeung2021learning} also employed a CNN to predict the position of 2D ultrasound fetal brain scans in 3D atlas space. 
However, the method only takes a set of images rather than image-volume pairs as input, which results in poor generalization ability of the model among individuals. 
For the ultrasound frame-to-volume registration, Guo~\etal~\cite{guo2021end} introduced an end-to-end registration network to align a 2D TRUS frame with a 3D TRUS volume. 
However, this method extracts features from ultrasound images only using 2D and 3D convolutions and directly concatenates them, which can be further enhanced by epicardium mask prompts to provide sufficient critical anatomical cues and adequate cross-dimensional feature interactions for the registration of ultrasound samples with low signal-to-noise ratios.

\begin{figure*}[t!]
	\centering
	\includegraphics[width=\linewidth]{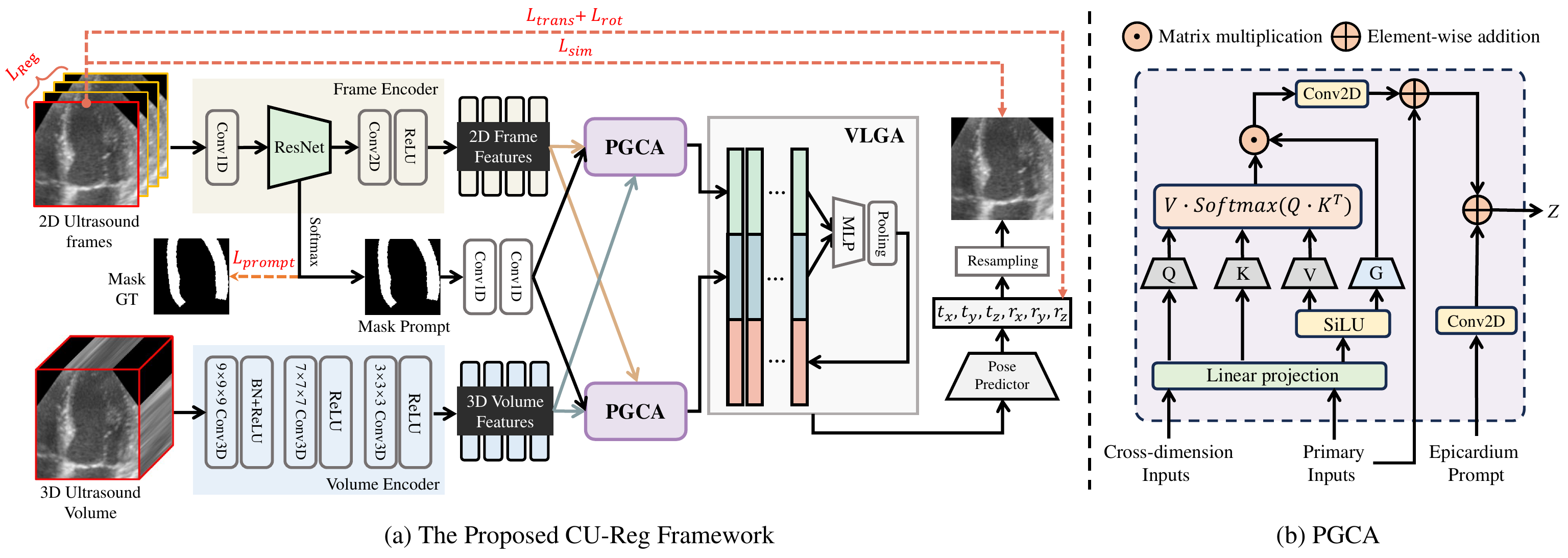}
	\caption{(a) Overview of the proposed~\ourmodel, where VLGA is the voxel-wise local-global aggregation; (b) The proposed prompt-guided gated cross-dimensional attention.}
	\label{fig:pipeline}
\end{figure*}

In this paper, we aim to accomplish a real-time and accurate cardiac ultrasound frame-to-volume registration to provide a complete guidance view for cardiac interventional surgery under the beating heart. 
To address the feature extraction difficulties caused by low signal-to-noise ratios and low tissue contrast in ultrasound images, we introduce epicardium mask prompts to provide sufficient critical anatomical information. 
Specifically, a bi-directional prompt-guided gated cross-dimensional attention (PGCA) operation is introduced to produce abundant structure features and perform efficient interaction between 2D frame and 3D volume features.
Further, we propose a voxel-wise local-global aggregation (VLGA) module to efficiently integrate dense local-global features across dimensions.
To avoid the large registration errors caused by small differences between adjacent frame images, we embed an inter-frame discriminative regularization term within our hybrid supervised learning to increase the distinction between adjacent slices in the same ultrasound volume to ensure registration stability. Additionally, we build a simulated cardiac frame-to-volume registration dataset through post-processing the CAMUS dataset~\cite{leclerc2019deep}.
The experimental results demonstrate that our model achieves superior performance compared to the state-of-the-art methods,
\eg, a runtime of over 35 FPS and a DistErr of 3.91 mm, which can meet the 5 mm accuracy requirement for many cardiac catheterizations~\cite{king2010respiratory}.
We hope our model can be applied for real-time and accurate guidance in cardiac interventions.

\section{Method}

\subsection{Overview of the proposed CU-Reg}

\figref{fig:pipeline}(a) illustrates the proposed lightweight end-to-end cardiac frame-to-volume registration network, called CU-Reg.
We consider real-time 2D ultrasound frame images and 3D ultrasound volumes as fixed and moving images respectively~\cite{guo2021end}, and take them as inputs to our framework.
CU-Reg outputs six parameters \{$t_x, t_y, t_z, r_x, r_y, r_z$\} that uniquely determine the spatial transformation of the 2D image coordinate system relative to the 3D volumetric coordinate system. 
Therein, the first three parameters determine the relative displacement between the origin of the coordinate system, and the last three parameters determine the rotation transformation matrix.

Given the moving image $U_v\in\mathbb{R}^{D\times H\times W}$ and fixed images $U_f^{i}\in\mathbb{R}^{H\times W},i=\{0,1,2,3\}$, where $U_f^{0}$ denotes the current anchor frame to be estimated and $U_f^{1, 2,3}$ denotes adjacent frames within the same volume.
We initially utilize two independent encoding branches to extract 2D slice features $\mathcal{F}_s$ and 3D volume features $\mathcal{F}_v$.
For the 2D frame branch, we first use two $Conv1d$ layers to normalize the channel dimension by increasing the number of channels from 4 to 64 followed by reducing it to 3, then feed frames into the CNN-based encoder~\cite{zhao2017pyramid} for multi-level features.
To encode the 3D volume, we employ three 3D convolutional blocks with different kernel sizes to extract coarse-to-fine multi-scale features.
Subsequently, we introduced the epicardium prompt-guided cross-dimensional attention operation that leverages the epicardium mask prompt with the bi-directional gated cross-dimensional attention block to spotlight critical anatomical features, providing informative alignment cues for ultrasound images. 
The enhanced features are processed by our voxel-wise local-global aggregation module to boost the fine-grained fusion of cross-dimensional representations.
Finally, transformation parameters are estimated via the pose predictor.
Additionally, our model embeds an inter-frame discriminative regularization term to highlight the discrimination between adjacent slices within the same ultrasound volume, yielding a hybrid-supervised training strategy to ensure registration stability.

\subsection{Epicardium Prompt-guided Cross-dimensional Interaction}

Due to the low contrast and signal-to-noise ratio of cardiac ultrasound slices, relying solely on the encoder is insufficient to provide critical anatomical information for intraoperative and preoperative registration.
In this regard, we exploit epicardium masks as prompts to pinpoint tissue landmarks for better alignment.
Specifically, the features $\mathcal{F}_s$ extracted from the 2D image encoder are passed through a $Softmax$ layer to epicardium mask prompts.
The predicted epicardium mask is supervised by the ready-made ground truth during training.
After passing the epicardium mask prompt through two 1$\times$1 convolutions for matching the dimensions of 3D volumetric features, we introduce a prompt-guided gated cross-dimensional attention (PGCA) to improve the interaction among 2D slice features, 3D volume features, and epicardium prompt features.
Inspired by gated attention~\cite{hua2022transformer}, PGCA dynamically regulates the feature dependencies between features of different dimensions, thereby enabling more efficient cross-dimensional interactions for capturing local-global features.
Here, we embed bi-directional PGCA operations, and the three inputs of each PGCA are the cross-dimensional input $\mathbf{C}\in \mathbb{R}^{d\times DHW}$, the primary input $\mathbf{P}\in \mathbb{R}^{d\times DHW}$, and the epicardium prompt $\mathbf{E}\in \mathbb{R}^{d\times DHW}$, where $\mathbf{P}$ represents the current branch features ($\mathcal{F}_s$ or $\mathcal{F}_v$), and $\mathbf{C}$ means other corresponding branching features ($\mathcal{F}_v$ or $\mathcal{F}_s$). 
In addition, $d$ and $DHW$ denote the feature channel dimension and the size of each feature map, respectively.
As described in~\figref{fig:pipeline}(b), we first perform a linear projection of $\mathbf{P}\in \mathbb{R}^{d\times DHW}$ and $\mathbf{C}\in \mathbb{R}^{d\times DHW}$ with the $SiLU$ function to produce the queries $Q$, keys $K$, values $V$ and gated vectors $G$:
\begin{equation}
    Q=W_q\cdot \mathbf{C}, \  K=W_k\cdot \mathbf{P}, \  V=\phi(W_v\cdot \mathbf{P}), \  G=\phi(W_g\cdot \mathbf{P}),
\end{equation}
where $W_q,W_k,W_v,W_g\in \mathbb{R}^{d\times d}$ denote the projection matrix, $\phi$ is the $SiLU$ function. 
Then, we obtain the enhanced 2D slice feature $z_{s}\in \mathbb{R}^{d\times DHW}$ and 3D volume feature $z_{v}\in \mathbb{R}^{d\times DHW }$ via the prompt-guided cross-attention, which can be formulated as follows:
\begin{equation}
    z_{i\in\{s,v\}}= \mathbf{P} + f(G\cdot \theta(Q\cdot K^{T} / \sqrt{d_{k}})\cdot V) \
    + f(\mathbf{E}),
\end{equation}
where $f(\cdot)$ denotes the convolution operations, $\theta(\cdot)$ is the standard $Softmax$ function, $1/\sqrt{d_{k}}$ is a scaling factor and $d_k$ is the number of channels.

\subsection{Voxel-wise Dense Local-Global Aggregation}

After obtaining the sufficient interaction between ultrasound frames and volume features, it is essential to fuse cross-dimensional features for a cohesive synthesis of critical structural details.
To accommodate ultrasound registration with multiple noises, we introduce a voxel-wise local-global aggregation module (VLGA) to efficiently associate local dense cues with global geometric information.
Given the enhanced 2D slice features $z_{s}\in \mathbb{R}^{d\times DHW}$ and 3D volume features $z_{v}\in \mathbb{R}^{d\times DHW}$ derived from bi-directional PGCA operations, 
we map the volume feature of each voxel and its spatial corresponding slice feature to the same size through 3D convolution and 2D convolution operations to generate voxel-wise pairs of features.
Subsequently, these feature pairs are concatenated and fed into an MLP to obtain a global feature vector $\mathcal{F}_{glo}$.
Lastly, $\mathcal{F}_{glo}$ is concatenated with the paired features, facilitating the acquisition of local-global context insights.
Our VLGA module can be summarized as follows:
\begin{equation}
    Z =\mathcal{C}[\mathcal{C}[z_s; z_v]; Max(MLP(\mathcal{C}[z_s; z_v))],
\end{equation}
where $\mathcal{C}[;]$ is the concatenation operation and $Max(\cdot)$ denotes max-pooling.

\subsection{Hybrid Supervised Learning}

In the training phase, we employ a hybrid loss function to supervise our model. 
Unlike existing methods that jointly regress the pose parameters, we propose to predict the pose parameters separately by decoupling the rotation and translation branches so as to avoid discontinuity in the rotational space from disturbing the prediction of translation parameters.
Here, we use two MLP layers to regress the rotation and translation parameters along with a smoothed L1 loss to supervise the pose parameters. 
The translation loss $\mathcal{L}_{trans}$ and rotation loss $\mathcal{L}_{rot}$ are used to make the network converge quickly, and they can be formulated as
\begin{equation}
    \mathcal{L}_{trans}=\frac{1}{N}\sum_{i=1}^{N}smooth_{L1}(\mathbf{T} - \mathbf{T}^*), \ 
    \mathcal{L}_{rot}=\frac{1}{N}\sum_{i=1}^{N}smooth_{L1}(\mathbf{R} - \mathbf{R}^*),
\end{equation}
where $N$ is the total number of samples. $\mathbf{T}$ and $\mathbf{R}$ means the predicted poses parameters, $\mathbf{T}^*$ and $\mathbf{R}^*$ are the ground truth.
For prompt learning, we utilize the MSE loss to directly supervise the predicted epicardium prompt $\mathbf{E}$: 
\begin{equation}
    \mathcal{L}_{prompt} = \frac{1}{N}\sum_{i=1}^{N}||\mathbf{E} - \mathbf{E}^*||_2,
\end{equation}
where $\mathbf{E}^*$ is the ground truth of mask prompts.
Moreover, to enlarge the discrimination between neighboring slices in the same volume, we embed an inter-frame discriminative regularization term $\mathcal{L}_{reg}$ to ensure the stability of the registration:
\begin{equation}
    \mathcal{L}_{reg}=\frac{1}{N}\sum_{i=1}^{N}smooth_{L1}(D_f - D^*_f),
\end{equation}
where $D_f \in \mathbb{R}^{1\times3}$ is the estimated inter-frame distance, which is predicted from the aggregated feature $Z$ by a separated MLP network, and $D^*_f$ is the ground truth.
The inter-frame distance is defined as the Euclidean distance between the translation vectors of the two adjacent frames.
In addition, we leverage the MS-SSIM loss $\mathcal{L}_{sim}$ for self-supervised training by constraining the similarity between the resampled and input frames to make the training process more stable and avoid overfitting.
Overall, the hybrid loss function of~\ourmodel~is computed as
\begin{equation}
    \mathcal{L} = \lambda_1 \mathcal{L}_{trans} + \lambda_2 \mathcal{L}_{rot} + \lambda_3 \mathcal{L}_{prompt} + \lambda_4 \mathcal{L}_{reg} + \lambda_5 \mathcal{L}_{sim}
\end{equation}
where $\lambda_{n},n=1,2,...,5$ are the hyper-parameters.
In our implementation, we set $\lambda_1=\lambda_2=1.0$, $\lambda_3=\lambda_4=0.1$, and $\lambda_5=0.5$, respectively.

\subsection{Implementation Details}
Our model is trained by an Adam optimizer on a single RTX3090 GPU for 500 epochs with a batch size of 16.
The inputs of~\ourmodel~are ultrasound slices with the size of 128$\times$128$\times$1$\times$1 and corresponding volumes with the size of 128$\times$128$\times$32$\times$1.
In the 2D ultrasound frame branch, we employ a ResNet-34~\cite{he2016deep} as the 2D encoder.
In the training phase, the inputs are a current anchor frame and three adjacent frames, \ie, 128$\times$128$\times$1$\times$4.
During inference, we manually adjust the input dimension by repeating the number of channels four times.
The hyperparameter values are the optimal results obtained through ablation experiments. 

\section{Experiments}

\subsection{Dataset and Evaluation Metrics}
To evaluate the cardiac frame-to-volume registration network, a simulated dataset is generated by post-processing the public CAMUS dataset~\cite{leclerc2019deep}. 
CAMUS contains 2D echocardiographic sequences with two- and four-chamber views of 500 patients, along with the masks of the left ventricular epicardium, these 2D sequences are expressed as 3D volumes in Cartesian coordinates with a unique grid resolution using the same interpolation procedure.
For each original 3D volume, four transformations are generated by add random deviations to the identity transformation, the deviations of translation parameters of each transformation are within the range of 10 mm, and the deviations of the rotation parameters are within the range of 20 degrees. 
Based on these transformations, 2D slices and corresponding masks are sampled from the original volume. The sampled slices include 128$\times$128 pixels, and the pixel spacing is 0.62 mm$\times$0.62 mm. 
Meanwhile, a new volume is sampled from the original volume based on the identity transformation with a volume size of 128$\times$128$\times$32 and a voxel spacing of 0.62 mm$\times$0.62 mm$\times$0.62 mm.
In this way, we can obtain four volume-frame-mask pairs with true transformations for one original volume. 
All data are split at the patient level, with 3,600 volume-frame-mask pairs for training and 400 volume-frame-mask pairs for testing.

For evaluation, we adopt the distance error (DistErr) to represent the average distance of the center and four corners between the input slice and the predicted slice. Normalized cross-correlation (NCC) and structure similarity index measure (SSIM) are used as image similarity metrics.
In addition, the translation error (TE) denotes the L1 distance between the true and the predicted translation vectors [$t_x, t_y, t_z$], and the rotation error (RE) denotes the L1 distance between the true and the predicted rotation vectors [$r_x, r_y, r_z$].

\begin{table*}[t!]
\caption{
Comparative results and ablation analysis of our~\ourmodel~in terms of mean values of quantitative metrics. 
$\uparrow$ / $\downarrow$ indicates the higher/lower the score, the better. 
}
\centering
\scriptsize
\renewcommand{\arraystretch}{0.98}
\renewcommand{\tabcolsep}{0.6mm}
\begin{threeparttable}[c]
\begin{tabular}{@{}l c c c ccc c@{}}
\toprule 
\multirow{2}{*}{Methods} & \multirow{2}{*}{\makecell[c]{DistErr\\(mm)$\downarrow$}} & \multirow{2}{*}{\makecell[c]{Img\\-NCC(\%)$\uparrow$}} & \multirow{2}{*}{\makecell[c]{Img\\-SSIM(\%)$\uparrow$}} &  \multicolumn{3}{c}{Transformation paprameters} & \multirow{2}{*}{\makecell[c]{Run-time\\(FPS)$\uparrow$}}\\
\cmidrule(r){5-7} 
    & & &  & TE(mm)$\downarrow$ & RE($^\circ$)$\downarrow$ & \makecell[c]{Para\\-NCC(\%)$\uparrow$}  & \\
\midrule
MRF-based \cite{porchetto2017rigid} & 4.02 & 87.14 & 60.06 & 2.51 & 6.49 & 72.83 & 0.1 \\
FVR-Net \cite{guo2021end} & 5.84 & 65.49 & 47.77 & 4.22 & 7.87 & 56.60 & 36 \cr\midrule
\textbf{CU-Reg} & \textbf{3.91} & \textbf{88.07} & \textbf{60.53} & \textbf{2.48} & \textbf{6.24} & \textbf{74.07} & 37 \\
w/o PGCA & 4.06 & 87.91 & 59.89 & 2.63 & 6.43 & 72.10 & 38 \\
w/o VLGA & 4.04 & 87.90  & 59.90 & 2.61 & 6.30 & 72.28 & 38 \\
w/o PGCA\&VLGA & 5.06 & 82.36 & 53.28 & 3.07 & 6.93 & 67.11 & \textbf{39} \\
w/o PGCA\&VLGA\&$\mathcal{L}_{reg}$ & 5.47 & 77.17 & 48.26 & 3.83 & 7.40 & 61.98 & \textbf{39} \\
Baseline & 7.99 & 60.39 & 44.39 & 4.81 & 8.54 & 49.72 & \textbf{39} \\
\bottomrule
\end{tabular}

\end{threeparttable}

\label{tab: Comparative result}
\end{table*}

\subsection{Comparison with State-of-the-Art methods}
\begin{figure*}[t]
    \centering
    \includegraphics[width=0.99\linewidth]{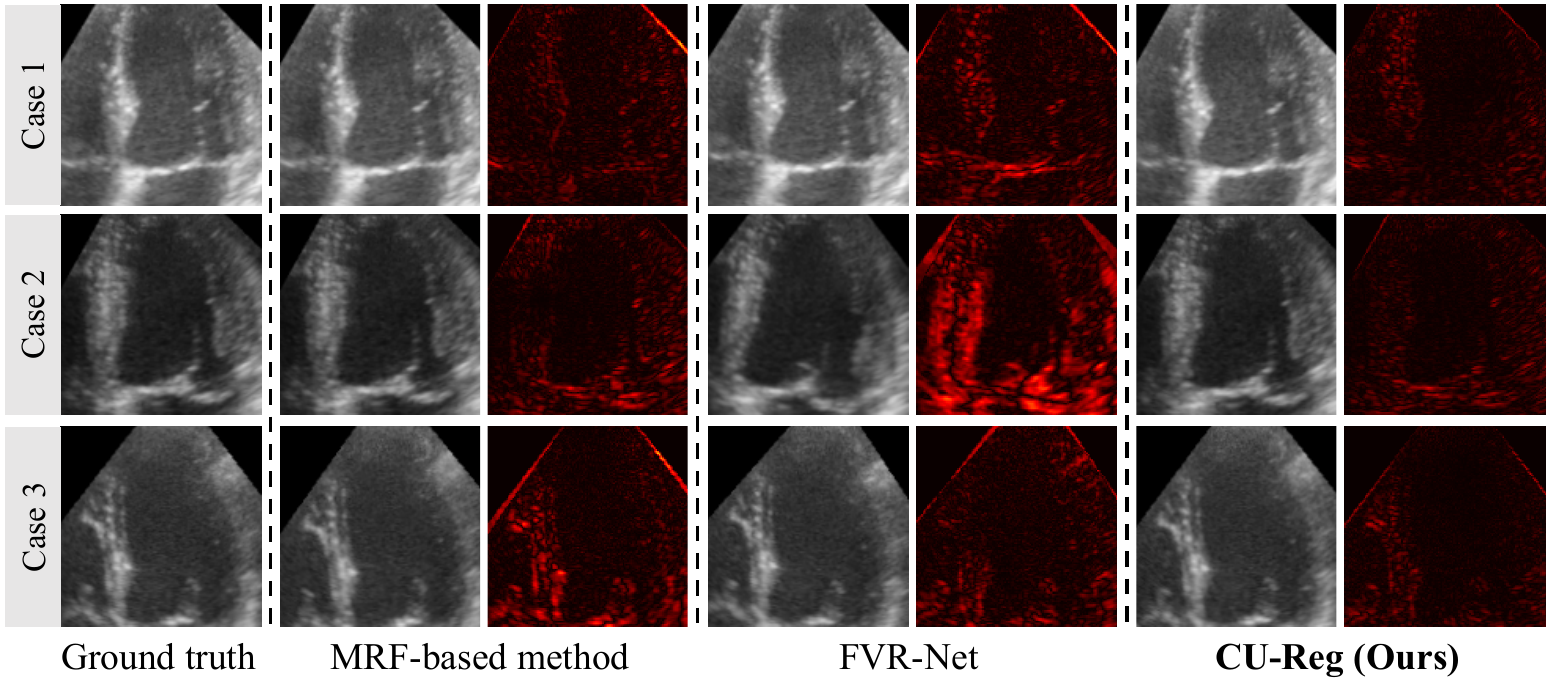}
    \caption{Qualitative comparison on the registration results of different methods, including predicted slices and their difference heatmaps with the ground truth.}
    \label{fig:vis_results}
\end{figure*}

We compare the proposed~\ourmodel~with the MRF-based conventional method~\cite{porchetto2017rigid} and the deep model FVR-Net~\cite{guo2021end} on the test set of our simulated data. 
As shown in \tabref{tab: Comparative result}, our model significantly outperforms FVR-Net for registration accuracy, \eg, about 33\% decrease for the DistErr and 34\% improvement for the Img-NCC, which can be attributed to the interaction of cross-dimensional features by the proposed PGCA and the augmentation of structural information by the epicardium mask prompt.
The visualization results in~\figref{fig:vis_results} also illustrate that our model can perform remarkable registration outcomes.
For registration efficiency, CU-Reg is significantly faster than conventional MRF-based methods (requiring multiple optimization iterations) by over 35 FPS, further confirming the superiority of our model in enhancing registration speed.

\textit{}

\subsection{Ablation Study}

We conduct thorough ablation experiments on each key component of the proposed~\ourmodel, including the epicardium prompt supervision, a prompt-guided gated cross-dimensional attention (PGCA), a voxel-wise local-global aggregation module (VLGA), and an inter-frame discriminative regularization term $\mathcal{L}_{reg}$. For our baseline, we utilize the regular 2D frame and 3D volume encoders to extract frame and volume features and directly concatenate them for feeding into the pose predictor.
As shown in the last two rows of~\tabref{tab: Comparative result}, when we embed the epicardium prompt supervision to the baseline, there is a significant improvement in the perception of anatomical features of cardiac ultrasound images by~\ourmodel, \eg, about 17\% increase in the Img-NCC metric.
With the addition of $\mathcal{L}_{reg}$ to our total loss function, the registration accuracy of our model is further improved.
Moreover, the proposed PGCA and VLGA play an indispensable role in the overall model and drive our model to optimal performance when used in synergy.
Additionally, the last column of ~\tabref{tab: Comparative result} illustrates the advantage of our model in inference speed, thanks to the lightweight design of~\ourmodel.

\section{Conclusion}

In this study, we present a novel lightweight end-to-end model, termed \ourmodel, for real-time and accurate cardiac ultrasound frame-to-volume registration. 
Launched from the epicardium mask prompt, we present a bi-directional prompt-guided gated cross-dimensional attention together with a voxel-wise local-global aggregation module to efficiently interact and integrate 2D sparse features and 3D dense features to obtain sufficient registration information.
Further, we also introduce inter-frame discriminative regularization to increase the discrimination of similar frames by our model. 
The experimental results demonstrate that the proposed~\ourmodel~outperforms the current state-of-the-art methods in both precision and efficiency.
Significantly, our model provides indispensable real-time guidance view for cardiac interventional surgery.
Furthermore, it can serve as a bridge for ultrasound-CT/MRI registration and showcase the potential for immediate application in cross-modal ultrasound-CT/MRI registration fields.

\begin{credits}
\subsubsection{\ackname} This work was supported in part by the Research Grants Council of the Hong Kong Special Administrative Region under Grant T45-401/22-N, in part by the Hong Kong Innovation and Technology Fund under Grants GHP/080/20SZ and GHP/050/20SZ, in part by the National Natural Science Foundation of China under Grants U23A20391 and 62273328, and in part by the Regional Joint Fund of Guangdong under Grants 2021B1515130003 and 2021B1515120011. 

\subsubsection{\discintname}
The authors have no competing interests to declare.
\end{credits}

\bibliographystyle{splncs04}
\bibliography{Paper-0300}
%





\end{document}